\documentclass[multphys,vecphys]{svmult}

\usepackage{makeidx}         
\usepackage{graphicx}        
\usepackage{multicol}        
\usepackage[bottom]{footmisc}

\usepackage{amssymb,amsmath}

\makeindex             

\newcommand{\figwidth}{\textwidth}
\newcommand{\medfigwidth}{0.7\textwidth}
\newcommand{\smallfigwidth}{0.5\textwidth}

\newcommand{\fref}[1]{Fig.~\ref{#1}}
\newcommand{\sref}[1]{Sec.~\ref{#1}}

\begin{document}

\title*{Dynamics of disordered elastic systems}
\author{T. Giamarchi\inst{1}\and
A. B. Kolton\inst{1}\and A. Rosso\inst{2}}
\institute{DPMC, University of Geneva, 24 Quai Ernest Ansermet, 1211
Geneva, Switzerland \texttt{Thierry.Giamarchi@physics.unige.ch}
\texttt{Alejandro.Kolton@physics.unige.ch} \and Laboratoire de Physique
Th\'{e}orique et Mod\`{e}les Statistiques B\^{a}t. 100, Universit\'{e}
Paris-Sud,
91405 Orsay Cedex, France \texttt{rosso@lptms.u-psud.fr}}
%
%
\maketitle
\section{Introduction}
Understanding the statics and dynamics of elastic systems in a
random environment is a longstanding problem with important
applications for a host of experimental systems. Such problems can
be split into two broad categories: (i) propagating interfaces such
as magnetic
\cite{lemerle_domainwall_creep,krusin_pinning_wall_magnet,repain_avalanches_magnetic,caysol_minibridge_domainwall}
or ferroelectric \cite{tybell_ferro_creep,paruch_2.5} domain walls,
fluid invasion in porous media \cite{wilkinson_invasion}, contact
line in wetting \cite{moulinet_contact_line}, epitaxial growth
\cite{barabasi_book} or crack propagation
\cite{bouchaud_crack_propagation}; (ii) periodic systems such as
vortex lattices
\cite{blatter_vortex_review,nattermann_vortex_review,giamarchi_vortex_review},
charge density waves \cite{gruner_revue_cdw}, or Wigner crystals of
electrons
\cite{andrei_wigner_2d,giamarchi_electronic_crystals_review}. In all
these systems the basic physical ingredients are identical: the
elastic forces tend to keep the structure ordered (flat for an
interface and periodic for lattices), whereas the impurities locally
promote the wandering. From the competition between disorder and
elasticity emerges a complicated energy landscape with many
metastable states. This results in glassy properties such as
hysteresis and history dependence of the static configuration.

To study the statics, the standard tools of statistical mechanics
could be applied, leading to a good understanding of the physical
properties. Scaling arguments and simplified models showed that even
in the limit of weak disorder, the equilibrium large scale
properties of disordered elastic systems are governed by the
presence of impurities. In particular, below four (internal)
dimensions, displacements grow unboundedly \cite{larkin_70} with the
distance, resulting in rough interfaces and loss of strict
translational order in periodic structures. To go beyond simple
scaling arguments and obtain a more detailed description of the
system is rather difficult and requires sophisticated approaches
such as replica theory \cite{mezard_variational_replica} or
functional renormalization group \cite{fisher_functional_rg}. Much
progress was recently accomplished both due to analytical and
numerical advances. For interfaces, the glassy nature of the system
is confirmed (so called {\it random manifold}), and a coherent
picture of the system is derived from the various methods. Periodic
systems have also been shown to have glassy properties but to belong
to a different universality class than interfaces, with quite
different behavior for the long range nature of the correlation
functions
\cite{giamarchi_book_young,nattermann_vortex_review,giamarchi_vortex_review}.

The competition between disorder and elasticity manifests also in the dynamics
of such systems, and if any in a more dramatic manner.  Among the dynamical
properties, the response of the system to an external force $F$ is specially
crucial, both from a theoretical point of view, but also in connection with
measurements. Indeed in most systems the velocity $v$ versus force $F$
characteristics is directly measurable and is simply linked to the transport
properties (voltage-current for vortices, current-voltage for CDW and Wigner
crystals, velocity-applied magnetic field for magnetic domain walls).

\begin{figure}
 \centerline{\includegraphics[width=\figwidth]{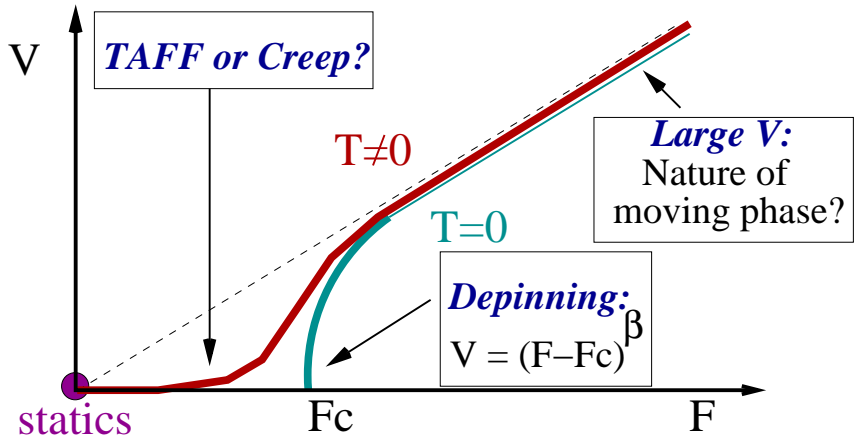}}
 \caption{\label{fig:vi} The velocity $v$ induced by an external force
 $F$ of a disordered elastic system. At zero temperature $T=0$ the
 system stays pinned until a critical force $F_c$ is reached. At
 finite temperature a motion can occur even for forces below
 threshold $F < F_c$ since the barriers to motion can always be
 passed by thermal activation.}
\end{figure}

Some of the questions related to this issue are shown in \fref{fig:vi}.
In the presence of disorder it is natural to expect that, at zero
temperature, the system remains pinned and only polarizes under the
action of a small applied force, i.e. moves until it locks on a
local minimum of the tilted energy landscape. At larger drive, the
system follows the force $F$ and acquires a non-zero asymptotic
velocity $v$. So a first set of questions is prompted by the zero
temperature properties. What is $F_c$ ? An estimate of $F_c$ can be
obtained via scaling arguments \cite{larkin_ovchinnikov_pinning} or
with a criterion for the breakdown of the large velocity expansion
\cite{schmidt_hauger,larkin_largev} and related to static quantities
such as the Larkin-Ovchinikov length, or computed numerically by an
exact algorithm \cite{rosso_depinning_simulation}. The $v-F$ curve
at $T=0$ is reminiscent of the one of an order parameter in a second
order phase transition \cite{fisher_depinning_meanfield}. Here the
system is out of equilibrium so no direct analogy is possible but
this suggests that one could expect $v \sim (F-F_c)^\beta$ with a
dynamical critical exponent $\beta$ \cite{duemmer_depinning}.
Calculation of such exponents is of course an important question
\cite{nattermann_stepanow_depinning,narayan_fisher_depinning,chauve_creep_long,rosso_hartmann}.

Another important set of questions pertain to the nature of the
moving phase itself, and in particular to the behavior at large
velocity: how much this moving system resembles or not the static
one
\cite{koshelev_dynamics,giamarchi_moving_prl,kolton_mglass_phases} ?
This concerns both the positional order properties and the
fluctuations in velocity such as the ones measured in noise
experiments.

Finally, one of the most important questions, and the one on which
we will concentrate in these notes, is the response well below
threshold $F \ll F_c$ at finite temperature. In this regime, the
system is expected to move through thermal activation. What is the
nature of this motion and what is the velocity? The simplest answer
would be that the system can overcome barriers via thermal
activation, \cite{anderson_kim} leading to a linear response at
small force of the form $v \sim e^{-U_b/T} F$, where $U_b$ is some
typical barrier. However it was realized that such a typical barrier
does not exist in a glassy system
\cite{nattermann_rfield_rbond,ioffe_creep,nattermann_creep_domainwall,feigelman_collective}
and that the response of the system was more complicated. The motion
is actually dominated by barriers which {\it diverge} as the drive
$F$ goes to zero, and thus the flow formula with finite barriers is
incorrect. Well below $F_c$, the barriers are very high and thus the
motion, usually called ``creep'' is extremely slow. Scaling
arguments, relying on strong assumptions such as the scaling of
energy barriers and the use of statics properties to describe an out
of equilibrium system, were used to infer the small $F$ response.
This led to a non linear response, characteristic of the creep
regime, of the form $v\sim \exp (- C\, F^{-\mu}/T)$.

Given the phenomenological aspect of these predictions and the
uncontrolled nature of the assumptions made, many open questions
remain to be answered, in particular whether such a behavior is
indeed correct \cite{kolton_string_creep} and can be derived
directly from microscopic equation of the motion
\cite{chauve_creep_short,chauve_creep_long}. We will review these
questions in these notes. The plan of the notes is as follows. In
\sref{sec:basic} we recall the basic concepts of interfaces in the
presence of disorder. In \sref{sec:creepphen}, we recall the
phenomenological derivation of the creep law. We present the
microscopic derivation of the creep law from the equations of motion
in \sref{sec:frg}, and discuss the similarities and differences with
the phenomenological result. In \sref{sec:wall} we focus on the
situation of domain walls. Such a situation is a particularly
important both for experimental realizations of the creep but also
because one dimension is the extreme case for such systems.
Conclusions can be found in \sref{sec:concl}.

\section{Basic concepts} \label{sec:basic}

\begin{figure}
 \centerline{\includegraphics[width=\medfigwidth]{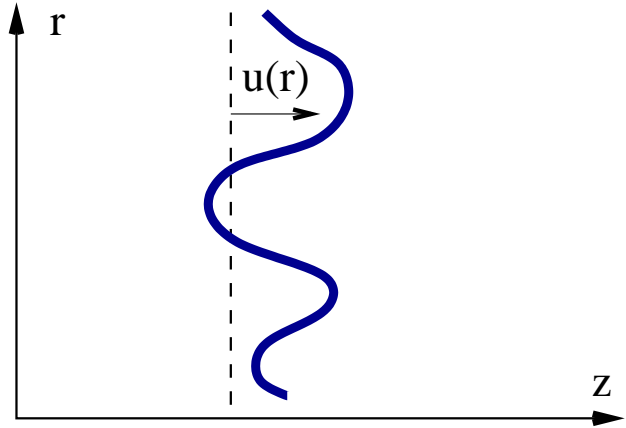}}
 \caption{\label{fig:elastic} A one dimensional interface (domain
   wall) living in a two dimensional space (film). The position of
   the interface is determined (provided there are no overhangs or bubbles) by
   the displacement $u$ from a flat configuration.}
\end{figure}

Let us introduce in this section the basic ingredients of the
systems under study. We will focus in these short notes to the case
of interfaces, but similar concepts apply to periodic systems as
well. The interface is a sheet of dimension $d$ living in a space of
dimensions $D$. For realistic interfaces $D = d+1$ but
generalization are of course possible (for example $D = d$
corresponds to periodic systems). We call $r$ the internal
coordinate of the interface and $z$ all its transverse directions.
The interface position is labelled by a displacement $u(r)$ from a
flat configuration. This determines totally the shape of the
interface provided that $u$ is univalued, i.e. that there are no
overhangs or bubbles. The case of a one dimensional interface
($d=1$) in a two dimensional film is shown in \fref{fig:elastic}.
Since the interface distortions cost elastic energy, its zero temperature
equilibrium configuration in the absence of disorder is the flat one.
Deviation from this equilibrium position are described by an Hamiltonian
$H[u]$ which is a function of the displacements $u$. For small displacements
one can make the usual elastic approximation
\begin{equation}
 H[u] = \frac12\int \frac{d^d q}{(2\pi)^d} c(q) u^*_q u_q
\end{equation}
where $u_q$ is the Fourier transform of $u(r)$ and $c(q)$ are the so called
elastic coefficients. If the elastic forces acting on the interface are short
ranged then one has $c(q) = c q^2$ which corresponds to
\begin{equation} \label{eq:elas}
 H[u] = \frac{c}{2} \int d^dr (\nabla u(r))^2
\end{equation}
For some interfaces where long range interactions play a role
different forms for the elasticity are possible. This is in
particular the case when dipolar forces \cite{nattermann_dipolar}
are taken into account \cite{paruch_2.5} or for the contact line in
wetting \cite{joanny_contact_line} and crack propagation
\cite{gao_crack}.

\begin{figure}
 \centerline{\includegraphics[width=\figwidth]{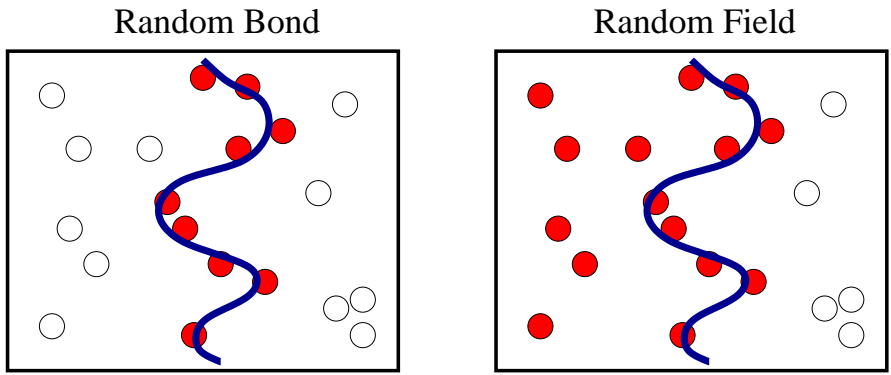}}
 \caption{\label{fig:rbrf} The two types of disorder (the names are
   coming from the magnetic realization of such systems). The dark circles are
   the impurities that contribute to the pinning energy of the interface.  In
   the random bond case only neighboring impurities contribute while in the
   random filed case all the impurities on the left side of the the interface
   contribute.  This makes the latter disorder effectively long ranged, even
   if the disorder potential $V(z,r)$ is short range.}
\end{figure}

In addition to the elastic energy the interface gains some energy by coupling
to the disorder. Two universality classes for the disorder exist (see
\fref{fig:rbrf}). The so called random bond disorder corresponds to impurities
that directly attract or repel the interface. On the contrary, for the so called random field
disorder the pinning energy is affected by all the randomness that the interface
has encountered in its previous motion. If $V(z,r)$ denotes the random
potential generated by the impurities the pinning energy writes:
\begin{equation}
   H_{\text{dis}}[u]       =     \int d r\begin{cases}
    V(u(r),r) & \text{random bond}\\
    \ \int_0^{u(r)} d z V(z,r)       & \text{random field}.
    \end{cases}
\label{e:disorder_class}
\end{equation}

The competition between disorder and elasticity manifests itself in
the static properties of the interface. The presence of disorder
leads to the appearance of many metastable states and glassy
properties. In particular, the interface deviates from the flat
configuration and becomes rough. From the scaling of the relative
displacements correlation function, a roughness exponent $\zeta$ can
be defined by
\begin{equation}
 B(r) = \overline{\langle [u(r)-u(0)]^2 \rangle} \propto r^{2\zeta}
\label{eq:displacements}
\end{equation}
where $\langle\ \ \rangle$ denotes thermodynamic average and
$\overline{\cdots}$ denotes disorder average. We will not enter in more
details about the statics here and refer the reader to the literature on that
point \cite{kardar_growth,barabasi_book}.

Dynamics is much more complicated since the standard tools of statistical
physics can not be used. One has to study the equation of motion of the system
\begin{equation} \label{eq:eqmotion}
\eta \frac{\partial}{\partial t}  u(r,t) = - \frac{\delta H}{\delta u(r,t)} + F +
\zeta(r,t)
\end{equation}
This equation is written for overdamped dynamics, but can include
inertia as well. $\eta$ is the friction taking into account the
dissipation, $F$ the external applied force, and $\zeta(r,t)$ a
thermal noise, needed to reproduce the effect of finite temperature.
The correlation of the thermal noise is $\langle
\zeta(r,t)\zeta(r',t')\rangle = 2 \eta T \delta(r-r')\delta(t-t')$.
Solving this equation of motion allows to extract all the dynamical
properties of the system. The presence of disorder in the
Hamiltonian $H$ makes this a very complicated proposal. In the
absence of the external force $F=0$, this Langevin equation allows
to recover the static properties after the system has achieved its
thermal equilibrium.

\section{Creep, phenomenology} \label{sec:creepphen}

\begin{figure}
\centerline{\includegraphics[width=\smallfigwidth]{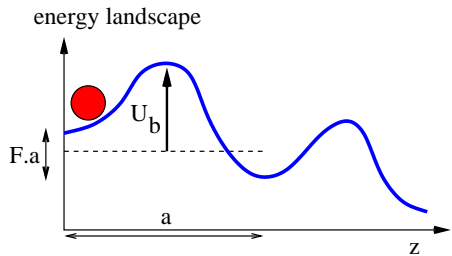}}
\caption{\label{fig:taff} In the Thermally Assisted Flux Flow (TAFF)
  \cite{anderson_kim} a region of pinned material is considered as a particle
  moving in an energy landscape characterized by  characteristic barriers $U_b$. This
  leads to an exponentionally small but linear response.}
\label{fig:landscape}
\end{figure}

Let us focus here on the response of the system to a very small
external force. For usual systems we expect the response to be
linear. Indeed earlier theories of such a motion found a linear
response. The idea is to consider that a blob of pinned material has
to move in an energy landscape with  characteristic barriers $U_b$ as shown in
\fref{fig:landscape}.
The external force $F$ tilts the energy landscape making forward
motion possible. The barriers are overcomed by thermal activation
(hence the name: Thermally Assisted Flux Flow) with an Arrhenius
law. If the minima are separated by a distance $a$ the velocity is
\begin{equation}
v \propto e^{-\beta(U_b - Fa/2)} - e^{-\beta(U_b + Fa/2)}
\simeq e^{- \beta U_b} F
\end{equation}
The response is thus linear, but exponentially small.

However this argument is grossly inadequate for a glassy system. The
reason is easy to understand if one remembers that the static system
is in a vitreous state. In such states a characteristic barrier
$U_b$ does not exist, since barriers are expected to diverge as one
gets closer to the ground state of the system.  The TAFF formula is
thus valid in systems where the glassy aspect is somehow killed and
the barriers do saturate. This could be the case for example for a
finite size interface. When the glassy nature of the system persists
up to arbitrarily large length scales the theory should be
accommodated to take into account the divergent barriers. This can
be done quantitatively within the framework of the elastic
description
\cite{nattermann_rfield_rbond,ioffe_creep,feigelman_collective,nattermann_pinning}.
The basic idea rests on two quite strong but reasonable assumptions:
(i) the motion is so slow that one can consider at each stage the
interface as motionless and use its {\it static} description; (ii)
the scaling for barriers, which is quite difficult to determine, is
the same as the scaling of the minimum of energy (metastable states)
that can be extracted again from the static calculation. If the
displacements scale as $u \sim L^\zeta$ then the energy of the
metastable states (see (\ref{eq:elas})) scales as
\begin{equation} \label{eq:scalmet}
E(L) \sim L^{d-2+2\zeta}
\end{equation}
where we use that elastic and pinning energy scale the same way. Since the
motion is very slow, the effect of the external force is just to tilt the
energy landscape
\begin{equation}
 E(L)  - F \int d^dr u(r) \sim L^{d-2+2\zeta}- F L^{d+\zeta}
\end{equation}
Thus, in order to make the motion to the next metastable state, one needs to
move a piece of the pinned system of size
\begin{equation} \label{eq:optimal}
L_{\text{opt} } \sim \left(\frac1F\right)^{\frac1{2-\zeta}}
\end{equation}
The size of the optimal nucleus able to move thus grows as the force decrease.
Since the barriers to overcome grow with the size of the object, the minimum
barrier to overcome ({\it assuming} that the scaling of the barriers is {\it
  also} given by (\ref{eq:scalmet}))
\begin{equation}
U_b(F) \sim \left(\frac1F\right)^{\frac{d-2 + 2\zeta}{2-\zeta}}
\end{equation}
leading to the well known {\it creep formula} for the velocity
\begin{equation}
 \label{eq:creep}
v \propto \exp \left[-\beta U_c\left(\frac{F_c}{F}\right)^{\mu}\right]
\end{equation}
where $F_c$ is the depinning force and $U_c$ a characteristic energy scale and
the {\it creep exponent} $\mu$ is given by,
\begin{equation}
\mu = {\frac{d-2 +2\zeta}{2-\zeta}}
\label{eq:creep_exponent}
\end{equation}
Equations (\ref{eq:creep}) and (\ref{eq:creep_exponent}) are quite
remarkable. They relate a dynamical property to {\it static}
exponents, and shows clearly the glassy nature of the system. The
corresponding motion has been called creep since it is a sub-linear
response. Of course the derivation given here is phenomenological,
and it will be important to check by other means whether the results
here hold. This will be the goals of the two next sections, where
first the creep law will be derived directly from the equation of
motion in $d=4-\epsilon$ dimensions, and then the creep will be
examined in the important case of $d=1$ domain walls.

\section{Around four dimensions} \label{sec:frg}

The previous phenomenological derivation of the creep formula rests on very
strong hypothesis. In particular it is assumed that: (a) the motion is
dominated by the typical barriers, and not by tails of distributions in the
waiting times or barriers; (b) the motion is so slow that the line has the
time to completely re-equilibrate between two hopping events so that one can
take all exponents as the equilibrium ones. Given the phenomenological ground
of these predictions and the uncontrolled nature of the assumptions made, both
for the creep and for the depinning, it is important to derive this behavior
in a systematic way, directly from the equation of motion.

In principle one has simply to solve the equation of motion
(\ref{eq:eqmotion}). In practice this is of course quite complicated. A
natural framework for computing perturbation theory in off-equilibrium systems
is the dynamical formalism \cite{janssen_dynamics_action,martin_siggia_rose}.
Integrating on all configurations $u$ we can exponentiate the equation
of motion by introducing an auxiliary field $\hat{u}$ :
\begin{multline}
 \int {\cal D}u \;\delta\left(\eta \frac{\partial u}{\partial t} +
\frac{\delta H}{\delta u(r,t)} - F - \zeta(r,t)\right) =\\
 \int {\cal D}u
 {\cal D}\hat{u} \;\exp\left[i \hat{u}(\eta \frac{\partial u}{\partial t}  +
\frac{\delta H}{\delta u(r,t)} - F - \zeta(r,t)) \right]
\end{multline}
the thermal and disorder average can easily be done, leading to a
field theory with some action $S$
\begin{multline}\label{actionuns}
 S(u,\hat{u}) = \int_{rt} i\hat{u}_{rt}(\eta \partial_t -
 c \nabla^2)u_{rt}
 -\eta T \int_{rt}i\hat{u}_{rt}i\hat{u}_{rt}
 -F\int_{rt}i\hat{u}_{rt}\\
 -\frac{1}{2} \int_{rtt'}
 i\hat{u}_{rt}i\hat{u}_{rt'}\Delta(u_{rt}-u_{rt'})
\end{multline}
where $\Delta$ is defined in the correlator of the pinning force $F_p=-\delta
H_{\text{dis}}/\delta u$, as
\begin{equation}
\overline{F_p(u,r) F_p(u',r') }= \Delta(u-u')\delta(r-r').
\end{equation}
The functional form of this correlator depends on whether one has random bond
or random field disorder (see e.g. \cite{chauve_creep_long} for more details).
Essentially $\Delta$ is a function with a width of the order of the
correlation of the disorder along the $z$ direction. The advantages of this
representation are many: disorder and thermal averages $\overline{\langle A[u]
  \rangle }=\langle A[u] \rangle_{S}$ of any observable $A[u]$ can be computed
with the weight $e^{-S}$; the response functions to an external perturbation
$h_{rt}$ are simply given by correlations with the response field: $\langle
A[u] i\hat {u}_{rt}\rangle = \frac{\delta}{\delta h_{rt}}\langle A[u] \rangle
$. In addition, since we are much more familiar with fields theories than with
equations of motion, we have at our disposal a variety of tools to tackle the
action $S$.
\begin{figure}
\centerline{\includegraphics[width=\medfigwidth]{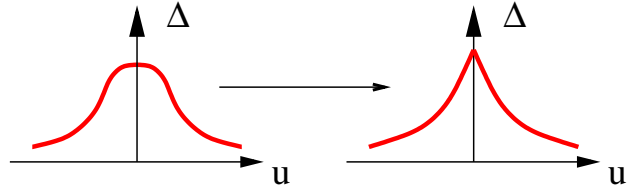}}
\caption{\label{fig:cuspfrg} Although the correlator of the disorder
  is initially an analytic function, a non analyticity (cusp) appears at a
  finite scale $l_c$. This length scale corresponds to the Larkin-Ovchinikov
  length at which pinning and metastability occur.}
\end{figure}
Although it is impossible to solve the action exactly it is possible
to look at its properties using a renormalization group procedure.
We will not detail the procedure but just recall here the resulting
functional renormalization group (FRG) flow equations
\cite{chauve_creep_short,chauve_creep_long} to give a flavor of
their physics
\begin{eqnarray}
 &&\partial \tilde{\Delta}(u)= (\epsilon-2\zeta ) \tilde{\Delta}(u)
 +\zeta u\tilde{\Delta}'(u)
 + \tilde{T} \tilde{\Delta}''(u) \label{flow}\\
 &&+ \int_{s>0,s'>0}
 \!\!\!\!\!\!\!\!\!\!\!\!\!\!\!\!\!\!\!\!e^{-s-s'}\left[
 \tilde{\Delta}''(u)\left(
 \tilde{\Delta}((s'-s)\lambda)-\tilde{\Delta}(u+(s'-s)\lambda)
 \right)\right. \nonumber \\
 &&-\tilde{\Delta}'(u-s'\lambda)\tilde{\Delta}'(u+s\lambda) \nonumber \\
 &&\left.+\tilde{\Delta}'((s'+s)\lambda)\left(
 \tilde{\Delta}'(u-s'\lambda)-\tilde{\Delta}'(u+s\lambda)\right)\right]
 \nonumber \\
 &&\partial \ln \lambda=
 2-\zeta -\int_{s>0}e^{-s}s \tilde{\Delta}''(s\lambda)\nonumber \\
 &&\partial \ln \tilde{T}= \epsilon-2-2\zeta +\int_{s>0}e^{-s}s
 \lambda \tilde{\Delta}'''(s\lambda)
 \nonumber \\
 &&\partial \tilde{F}= e^{ -(2-\zeta )l} c \Lambda_{0}^2 \int_{s>0}
 e^{-s} \tilde{\Delta}'(s\lambda) \nonumber
\end{eqnarray}
where $\epsilon =4-d$, $\partial$ denotes $\frac{\partial}{\partial
l}$ and $\lambda = \eta_l v$.  The tilde denotes rescaled quantities
(see \cite{chauve_creep_long} for the notations). Contrarily to the
standard case of critical phenomena, where the potential $\Delta$
can be expanded in powers of the field and only the first terms are
relevant, here all the powers of the expansion have the same
dimension. It is thus necessary to renormalize the whole function
\cite{fisher_frg_1} (in other words one has an infinite set of
coupled renormalization equations). One of the most important
consequences is shown in \fref{fig:cuspfrg}: the renormalized
function $\Delta$ becomes nonanalytic beyond a certain length scale,
and develop a cusp which signals pinning and the glassy properties
of the system.  This cusp appears at a finite length scale
corresponding to the Larkin-Ovchinikov length \cite{larkin_largev}
and it is directly related to the existence of the finite critical
force $F_c$
\cite{nattermann_stepanow_depinning,narayan_fisher_depinning} at
zero temperature. The FRG procedure has been push up to two loop
expansion \cite{2loop,wiese_frg_review}.

\begin{figure}
\centerline{\includegraphics[width=\figwidth]{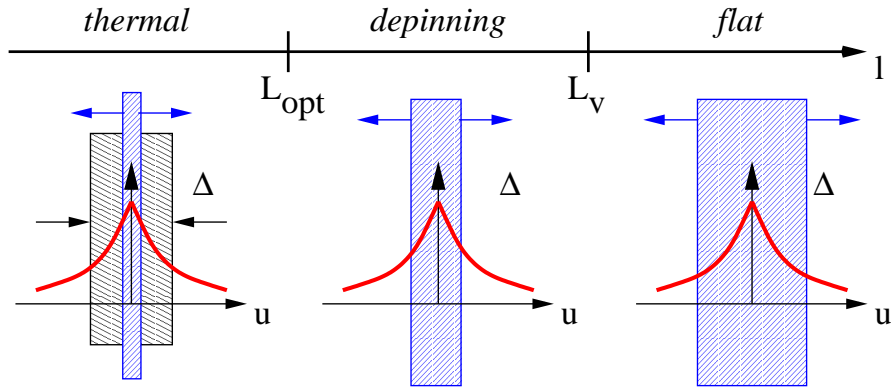}}
\caption{\label{fig:cusp} The cusp is rounded by both the
  temperature and the finite velocity of the interface. In the thermal regime,
  the main source of rounding comes from the temperature and the role of
  velocity is negligible. Then the system enters a regime in which the main
  source of rounding is the velocity and the role of temperature is
  negligible. This regime is very similar to a depinning regime. Finally the
  velocity rounds the whole correlator of the disorder and thus disorder is washed
  out by the averaging due to motion. At this length scale one recovers a
  purely thermal interface.}
\end{figure}

The presence of a finite temperature and a finite velocity
(proportional to $\lambda$) prevents the appearance of the cusp
\cite{chauve_creep_short,chauve_creep_long}. For very small external
force, the way the cusp is cut occurs in two steps, as shown in
\fref{fig:cusp}.
At very low velocity the cusp is cut first by the temperature and
the velocity can be forgotten. A physical way to interpret this
regime is that the motion consists essentially in overcoming the
barriers by thermal activation. This regime corresponds essentially
to the one in the phenomenological derivation of the creep.
Increasing the scale $l$ the temperature renormalizes down and the
velocity renormalizes up. For this reason, above a lengthscale that
can be identified with $L_{\text{opt}}$, the cusp starts to be
regularized by the velocity.  The temperature can now be forgotten
and a regime very similar to the standard depinning regime at $T=0$
takes place. Then, finally, at a certain length scale $L_v$ the
whole $u$ dependence of the correlator of the disorder $\Delta(u)$
is erased by the finite velocity. This corresponds to a regime where
the motion of the interface has averaged over the disorder and thus,
in the moving frame, the interface is now simply submitted to the
thermal-like noise \cite{nattermann_stepanow_depinning}.

\begin{figure}
 \centerline{\includegraphics[width=\figwidth]{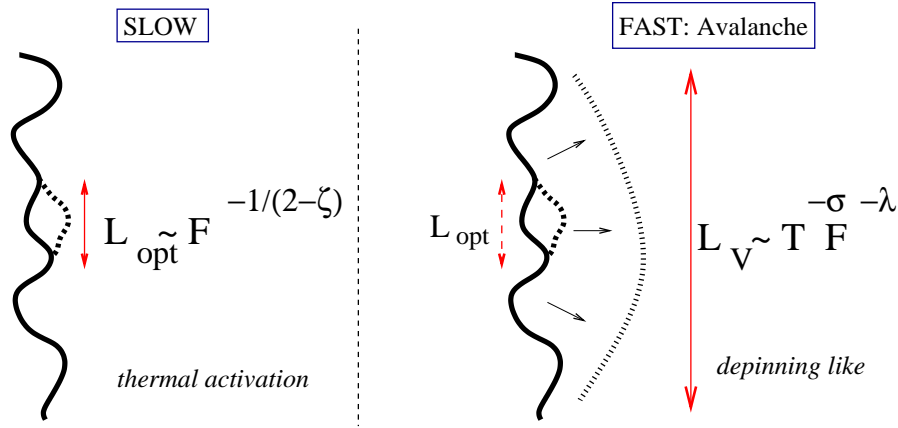}}
 \caption{\label{fig:aval} The phenomenological theory of creep (left) would
   predict that the avalanche size corresponds to the size of the thermal
   nucleus $L_{\text{opt}}$. From the FRG a quite different size emerges
   corresponding to a larger avalanche of size $L_v$ triggered by the thermal
   motion, in a way similar than for a depinning process.  $\lambda$ and
   $\sigma$ are two characteristic exponents.}
\end{figure}

The FRG calculation of the velocity confirms the phenomenological arguments
and finds the creep law (\ref{eq:creep}). Moreover, at the first order in
$\epsilon$, the creep exponent $\mu$ agrees with the scaling
(\ref{eq:creep_exponent}). The velocity behavior is thus dominated by the
first (thermal) regime. On the other hand because the second regime exists, we
expect that the phenomenological derivation is incorrect as far as the
characteristic lengthscales of the problem are concerned. Indeed, the
phenomenological derivation would predict that the characteristic size of a
moving domain is the optimal length scale $L_{\text{opt}}$, which coincides
with the end of the thermal regime. However the FRG equations predict that the
size of a moving domain corresponds to $L_v$, the scale of the end of the
depinning regime. A physical way to understand this behavior is to say that
the motion of the thermal nucleus, of size $L_{\text{opt}}$, triggers an
avalanche of a larger size $L_v$. The FRG thus predicts much larger avalanche
scales than what should be naively expected from the phenomenological theory,
as shown in \fref{fig:aval}.
Such large avalanches are in agreement with recent experiments in magnetic
systems \cite{repain_avalanches_magnetic}.

Confirming the stretched exponential behavior of the creep is of
course an experimental challenge given the large span of velocities
needed. The first unambiguous determination of the creep law with a
precise determination of the exponent was made in magnetic films,
for one dimensional domain walls \cite{lemerle_domainwall_creep},
and confirmed with subsequent measurements
\cite{repain_avalanches_magnetic,caysol_minibridge_domainwall}.
Ferroelectric systems \cite{tybell_ferro_creep,paruch_2.5} have
shown a creep exponent compatible with two dimensional domain walls
in presence of dipolar forces . In periodic systems, such as
vortices, it is more delicate to determine the precise value of the
exponent, even when non linear behavior has been clearly observed.
The experiment \cite{fuchs_creep_bglass} shows a creep exponent in
agreement with the theoretical predictions.
\begin{figure}
 \centerline{\includegraphics[width=7.0cm]{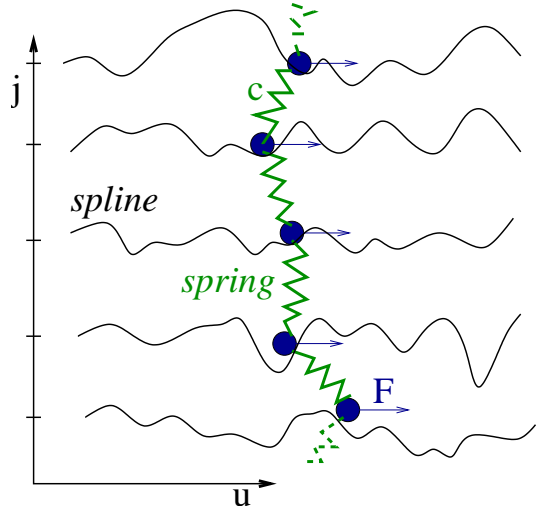}}
 \caption{Discretization scheme for the elastic line
 driven in a random potential.}
 \label{fig:geometry}
\end{figure}

\section{Low dimensional situation: domain walls} \label{sec:wall}

Around four dimensions the creep hypothesis gives a velocity
dependence consistent with the one obtained from the microscopic
derivation, at least up to the order $\epsilon$ at which the
renormalization group analysis can be performed. Let us now focus at
the other extreme limit, namely when the wall is one dimensional and
moves in a two dimensional space. The interest in such a situation
is twofold. First, as already mentioned, controlled experiments are
performed on domain wall motion. Second, from the theoretical point
of view the situations of a low dimensional domain wall is very
interesting. Thermal effects are increasingly important as the
dimension is lowered. For $d \leq 2$ they lead to a roughening of
the domain wall, even in the absence of disorder (with an exponent
$\zeta_T = (2-d)/2$). One can thus expect more intricate competition
between temperature and disorder effects.

Numerical simulations are a valuable alternative theoretical tool to
address this open issue. In this respect, Langevin dynamics
simulations have been used to study both the velocity-force
($v$-$F)$ characteristics and the dynamic roughness $\zeta$ of an
elastic string in a random potential
\cite{kaper_creep_simulation,chen_marchetti,kolton_string_creep}. In
\cite{kolton_string_creep} we have studied equation
(\ref{eq:eqmotion}) with a short range elasticity:
\begin{equation}
\eta \frac{\partial }{\partial t}u(r,t) = c \partial_r^2 u(r,t) + F_p(u,r) + F + \zeta(r,t)
\label{eq:eqmotion_num}
\end{equation}
where $F_p(u,r)=-\partial_u V(u(r),r)$ is the pinning force derived from the
random bond disorder $V(u,r)$.

\begin{figure}
 \centerline{\includegraphics[width=8.5cm]{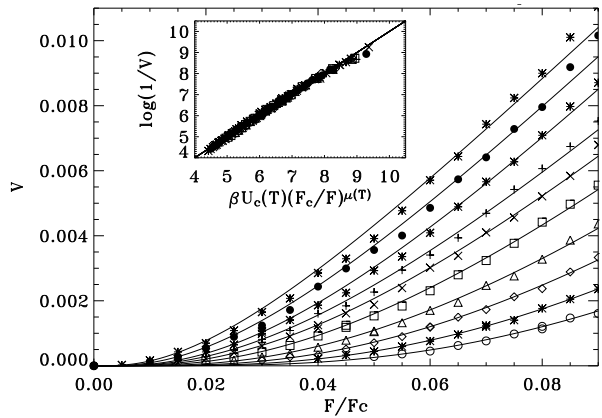}}
 \caption{After \cite{kolton_string_creep}. $v$-$F$ characteristics for several
   temperatures, increasing from bottom to top. Solid lines are fits of the
   creep formula (\ref{eq:creep}) with $U_c$ and $\mu$ as fitting parameters.
   The inset assures the validity of the creep formula in the range of
   temperature and velocity analyzed.}
 \label{fig:vf}
\end{figure}

To solve numerically (\ref{eq:eqmotion_num}) we discretize the
string along the $r$ direction, $r\rightarrow j=0,\ldots,L-1$,
keeping $u_j(t)$ as a continuous variable. A second order stochastic
Runge-Kutta method \cite{greenside_runge_kutta} is used to integrate
the resulting equations. To model a continuous random potential we
generate, for each $j$, a cubic spline $V(u_j,j)$ passing through
regularly spaced uncorrelated Gaussian random points
\cite{rosso_depinning_simulation}. The geometry of our system is
shown in \fref{fig:geometry}. We are interested in the $v$-$F$
characteristics.  Typical curves, obtained in the simulations, are
shown in \fref{fig:vf}.
In the whole range of temperature and pinning strength analyzed we find that
the $v$-$F$ curve can be well fitted by the creep formula (\ref{eq:creep})
with $U_c$ and $\mu$ as fitting parameters. We thus confirm the predicted
stretched exponential behavior. However, contrarily to the naive creep
relation (\ref{eq:creep_exponent}) we find that not only $U_c$, but also
$\mu$, depend on temperature. Since the phenomenological theory assumes that
$\mu$ can be computed directly from the roughness exponent $\zeta$ it is
important to study the geometrical properties of the driven string. For this
reason we introduce the averaged structure factor,
\begin{eqnarray}
S(q) \sim \overline{\biggl\langle \biggl|\int dr u(r,t) e^{-iqr}\biggr|^2  \biggr\rangle}.
\label{S}
\end{eqnarray}
\begin{figure}
 \centerline{\includegraphics[width=8.5cm]{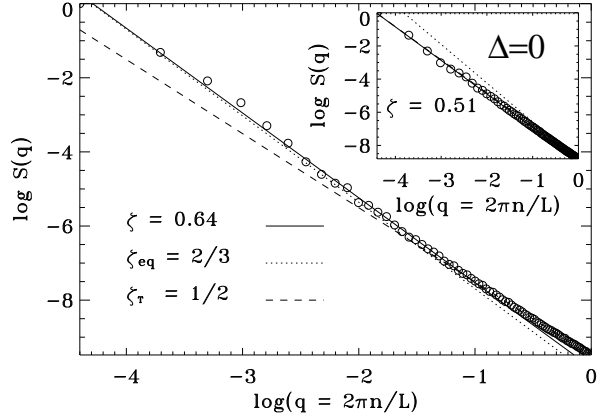}}
 \caption{Structure factor $S(q)$ of the elastic line for the statics.
   In the presence of disorder the long distance behavior is characterized by
   a roughness exponent $\zeta_{eq}=2/3$, while in the absence of disorder
   (see inset) thermal fluctuation gives $\zeta_T=1/2$.}
 \label{fig:statics}
\end{figure}
The dimensional analysis of this double integral allow us to compute $\zeta$
from $S(q) \sim q^{-(1+2\zeta)}$, valid for small $q$. In \fref{fig:statics}
we show the structure factor of an elastic string thermally equilibrated at
$F=0$. We can observe a crossover between a short distance regime where
thermal fluctuations are dominant ($\zeta \sim \zeta_T =1/2$) and a long
distance disorder dominated regime where we find the well known roughness
exponent $\zeta \sim \zeta_{eq}=2/3$ \cite{kardar_exponent_line}.
\begin{figure}
 \centerline{\includegraphics[width=8.5cm]{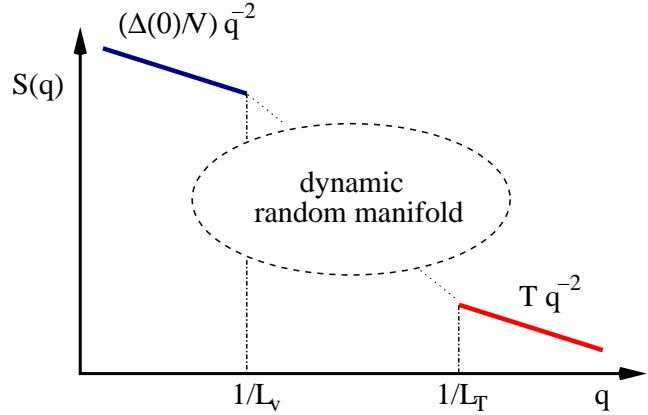}}
 \caption{Sketch of $S(q)$ expected for a driven elastic line.}
 \label{fig:dream}
\end{figure}
For the dynamics (see \fref{fig:dream}), when $F\neq 0$ one can
predict that the short distance behavior $L < L_T$ of the elastic
string is still dominated by thermal fluctuations ($\zeta_T=1/2$).
Note that this thermally dominated regime has nothing to do with the
regime derived in the previous section and valid up to the scale
$L_{\text{opt}}$. In this regime disorder is already dominant and
barriers are overcomed by thermal activation.  On the other hand, as
already discussed, the finite velocity makes the quenched disorder
to act as a thermal noise at the largest length scale $l>L_v$. Thus,
in this case, the expected exponent is also $\zeta_v=1/2$
\cite{nattermann_stepanow_depinning}.  Finally, at intermediate
length scales, the physics is determined by the competition between
disorder and elasticity and characterized by a non trivial random
manifold roughness exponent.

\begin{figure}
 \centerline{\includegraphics[width=10cm]{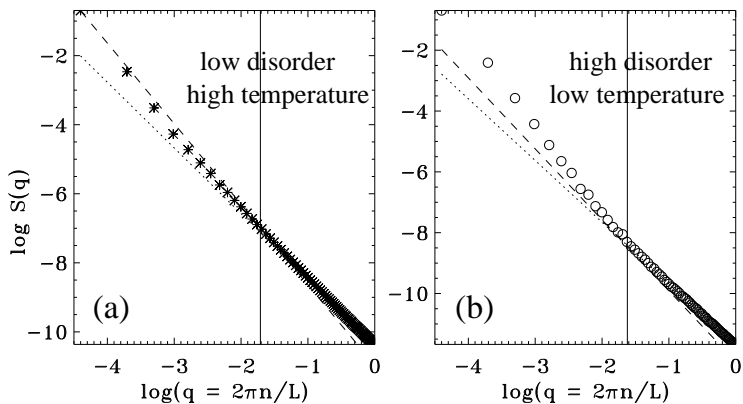}}
\caption{After \cite{kolton_string_creep}. Structure factor $S(q)$
of the
  elastic line in the driven case. Two regimes are distinguished. (a) In the
  low disorder/high temperature regime the roughness exponent is consistent
  with $\zeta_{eq}=2/3$. (b) In the high disorder/low temperature regime the
  roughness exponent is clearly bigger than $\zeta_{eq}$.}
\label{fig:roughness}
\end{figure}

\begin{figure}
 \centerline{\includegraphics[width=10cm]{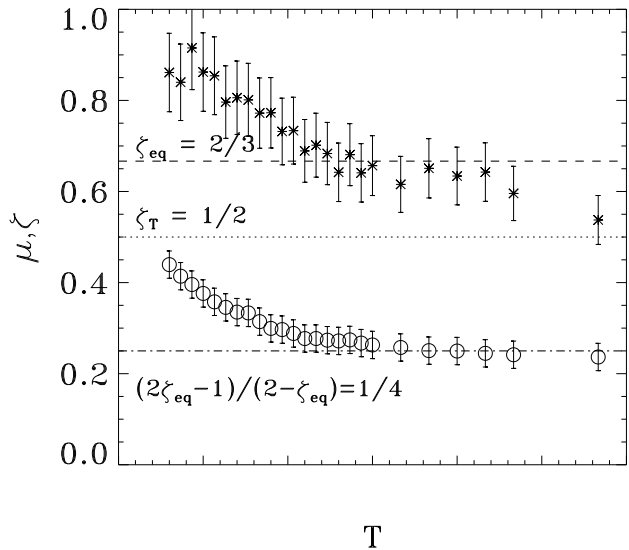}}
\caption{After \cite{kolton_string_creep}. Roughness exponent,
$\zeta(T)$, and
  creep exponent, $\mu(T)$, vs $T$. The dashed line gives the equilibrium
  roughness exponent $\zeta_{eq}=2/3$, and the dotted line the purely thermal
  roughness $\zeta_T=1/2$. The expected value for the creep exponent $\mu=1/4$
  is also indicated.}
\label{fig:exponents}
\end{figure}

A systematic analysis of the $v$-$F$ characteristics and $S(q)$ show
essentially two different regimes of creep motion. In
\fref{fig:roughness}(a) and (b) we show the structure factor for the
two cases.
As predicted, we get $\zeta \sim \zeta_{T}=1/2$ for large $q$. At a
certain scale we observe a crossover between the thermal and the
random manifold scaling. The location of this crossover decreases as
temperature (disorder) is increased (decreased). We can also observe
that the second velocity-controlled crossover is not achieved in our
finite-size simulation due to the very slow dynamics. Interestingly,
for the small disorder case (\fref{fig:roughness}(a)) the random
manifold scaling gives $\zeta = 0.67 \pm 0.05$, in excellent
agreement with the equilibrium value $\zeta_{eq}=2/3$, while a much
higher roughness exponent $\zeta = 0.9 \pm 0.05$ is found for the
strong disorder case (\fref{fig:roughness}(b)).  The analysis of the
$v$-$F$ characteristics brings us to the same conclusion: the value
of the exponent $\mu$ is close to the equilibrium value $\mu=1/4$
for low disorder (high temperature) and departs from this value when
the disorder (temperature) increases (decreases). In
\fref{fig:exponents} we summarize all the results.
We notice that although the values of $\zeta$ and $\mu$ depart from
the equilibrium values, the relation (\ref{eq:creep_exponent}) seems
still to hold, within the error bars for the two exponents. This is
highly non-trivial since equation (\ref{eq:creep_exponent}) is
derived from a calculation of the barriers in an equilibrium
situation.

We thus have found two regimes of creep motion.  The first one
occurs when the temperature is larger than the strength of the
disorder, giving $\mu \sim 1/4$ and $\zeta \sim 2/3$ as predicted by
assuming a quasi-equilibrium nucleation picture of the creep motion.
This implies that the domain wall has time to re-equilibrate between
hops, being the underlying assumption behind
(\ref{eq:creep_exponent}) essentially satisfied.  The second regime
occurs for temperatures smaller than the strength of the disorder,
and is characterized by anomalously large values of both exponents.
This clearly shows that in this regime the domain wall stays out of
equilibrium, and that the naive creep hypothesis does not apply.
Note that the measured roughness exponent is intermediate between
the equilibrium value and the depinning value $\zeta_{dep} = 1.26
\pm 0.01$ \cite{rosso_depinning_simulation}. The fact that the
thermal nucleation which is the limiting process in the creep
velocity, is in fact followed by depinning like avalanches was noted
in the FRG study of the creep \cite{chauve_creep_long}.  Whether
such avalanches and the time it would take them to relax to
equilibrium is at the root of the observed increase of the exponent,
is clearly an interesting but quite complicated open question.

\section{Conclusions and open questions} \label{sec:concl}

In these short notes we have presented a brief review of the dynamical
properties of interfaces in a disordered environment. We have in particular
focused on the response of such interfaces to a very small external force,
and the corresponding very slow motion it entails (so called creep). Clearly
many important questions remain to be understood for this problem. for large
dimensions the microscopic derivation clearly supports the phenomenological
one as far as the velocity is concerned, but also shows that different
length scales enter to describe the dynamics. In particular, it predicts a much
larger avalanche size than initially anticipated.  For one dimensional walls
the situation is even more complex, and the very hypothesis that the wall is
constantly in equilibrium between two creep processes seems incorrect at least
when the disorder is not weak enough or if the temperature becomes too low.
The deviations such effect might entail on the creep exponent is of course
important in connection with the experimental work.

Of course these questions are only the tip of the iceberg and more
subtle questions such as how such a domain wall can age in the
presence of the disorder are still largely not understood, and more
analytical, numerical or experimental work is clearly needed to
address these issues.

\section{Acknowledgments}

We have benefitted from invaluable discussions with many colleagues,
too numerous to thank them all here. We would however like to
specially thank D. Dom\'{\i}nguez, J. Ferr\'e, J. P. Jamet, W.
Krauth, S. Lemerle, P. Paruch, V. Repain,  J.M. Triscone.  TG
acknowledges the many fruitful and enjoyable collaborations with P.
Le Doussal and P. Chauve. This work was supported in part by the
Swiss National Science Foundation under Division II.



\printindex
\end{document}